\documentclass{article}

\usepackage{arxiv}

\usepackage[utf8]{inputenc}         
\usepackage[T1]{fontenc}            
\usepackage[colorlinks]{hyperref}   
\usepackage{url}                    
\usepackage{booktabs}               
\usepackage{amsfonts}               
\usepackage{nicefrac}               
\usepackage{microtype}              
\usepackage{graphicx, caption}

\hypersetup{
    colorlinks=true,
    citecolor=magenta,
    linkcolor=blue,
    filecolor=magenta,      
    urlcolor=blue,
    }
    
\title{Introducing the Voxel Interactive Contour Tool for Online Radiation Intensity Analytics (VICTORIA)}

\author{
  Elise Badun \\
  University of Victoria\\
  Victoria, BC, Canada\\
  \texttt{ebadun@uvic.ca}
  
 \And    
  Fr\'ed\'eric Tessier \\
  National Research Council Canada\\
  Ottawa, ON, Canada\\
  \texttt{Frederic.Tessier@nrc-cnrc.gc.ca}
  
  \And
  Reid Townson \\
  National Research Council Canada\\
  Ottawa, ON, Canada\\
  \texttt{Reid.Townson@nrc-cnrc.gc.ca}
  
  \And
  Ernesto Mainegra-Hing \\
  National Research Council Canada\\
  Ottawa, ON, Canada\\
  \texttt{Ernesto.Mainegra-Hing@nrc-cnrc.gc.ca}
  
  \And
  Margaret-Anne Storey \\
  University of Victoria\\
  Victoria, BC, Canada\\
  \texttt{mstorey@uvic.ca}
  
     \And
 Magdalena Bazalova-Carter \\
  University of Victoria\\
  Victoria, BC, Canada\\
  \texttt{bazalova@uvic.ca}
  
}

\begin{document}
\maketitle

\begin{abstract}
In this paper, the Voxel Interactive Contour Tool for Online Radiation Intensity Analytics (VICTORIA) web viewer is presented as a solution to the issue of inaccessible, expensive, and insecure Digital Imaging and Communications in Medicine (DICOM) and Monte Carlo (MC) dose viewers. VICTORIA is a tool for the visualization and comparison of dose distributions with underlying anatomy that is accessible, free, designed with security in mind, and available on the web\footnote{\url{http://web.uvic.ca/~bazalova/VICTORIA/}}. The code is open-source and under the GNU General Public License, and it can be found at the EGSnrc Github repo in the DICOM-viewer branch\footnote{\url{https://github.com/ellieb/EGSnrc/tree/DICOM-viewer/HEN_HOUSE/gui/dose-viewer}}. The viewer should be useful for researchers using EGSnrc file types \textit{.egsphant} and \textit{.3ddose} from the EGSnrc toolkit, patients who want a simple tool to view files, or researchers using DICOM Computed Tomography (CT) and DICOM Radiotherapy (RT) Dose files in low- or middle-income countries without access to treatment planning systems.
\end{abstract}

\section{Introduction}
\label{sec:introduction}

Many popular software tools exist to visualize computed tomography (CT) images. Programs like Osirix, 3D Slicer, ImageJ, Dicompyler, and Horos are used to analyze and compare medical images. As important as these tools are, almost none of them (with the exception of ImageJ) are free and browser-based. If using software-specific formats such as \textit{.egsphant} and \textit{.3ddose} files, online visualization programs are challenging to find. The goal of VICTORIA is to be an easy tool to view and compare different dose and density files in a standard web browser, requiring no additional software download and installation. VICTORIA was initially developed as a web tool that could easily view and compare phantom and dose files from the EGSnrc toolkit, an Electron Gamma Shower software package by the National Research Council of Canada (NRC). The EGSnrc Monte Carlo (MC) toolkit is a well-known medical physics package that simulates the transport of ionizing radiation through matter \cite{egsnrc_2021}. It can model the transport of electrons, positrons, and photons with energies ranging from 1~keV to 10~GeV, in user-defined geometries. The EGSnrc package uses custom file types \textit{.egsphant} and \textit{.3ddose}, which are formatted plain text files containing phantom and dose information.

The toolkit comes with legacy tools to visualize these file types however, they are outdated and can be hard to install on some operating systems. For example, a user who wants to run the EGSnrc \textit{dosxyz\_show} program on a Windows computer would have to install a virtual machine to run the program on Linux. These types of restrictions prompted the collaboration between the NRC and the University of Victoria to build a web-based \textit{.egsphant} and \textit{.3ddose} viewer. Work began on the project under the supervision of Magdalena Bazalova-Carter in conjunction with Frédéric Tessier and Reid Townson and Ernesto Mainegra-Hing from the NRC. Later, in collaboration with Margaret-Anne Storey, the ability to load and view DICOM files was added, as requested by users of the site.

\section{Materials and Methods}
\label{sec:materials-and-methods}
VICTORIA is a web-based viewer coded in JavaScript, HTML, and CSS. Most of the data visualization capabilities come from D3, a JavaScript library made for producing dynamic data-based displays by binding input to the Document Object Model (DOM) and applying transformations \cite{bostock_2020}. D3 provides many user interface functions in VICTORIA, particularly data plotting, axes support, click events, zooming, and other changes to the DOM. The dicomParser library from Cornerstone is used to read in and parse DICOM files \cite{dicomparser_2020}. The other libraries in use are d3-legend, which creates colour-based legends, as well as filesaver and dom-to-image, which assists in saving the volume viewer to PNG \cite{d3legend_2019, filesaver_2020, domtoimage_2017}. The code for this project can be found in the EGSnrc GitHub repository \footnote{\url{https://github.com/ellieb/EGSnrc/tree/DICOM-viewer/HEN_HOUSE/gui/dose-viewer}}.

Due to the sensitive nature of medical imaging documents, all processing is done by the browser. This means that any data loaded into the viewer remains in the client context, so no files are uploaded over a network to a database, improving file security. To improve site performance, Web Workers are used to cache density slices. Web Workers allow for code execution in the background without interfering with the user interface. Since VICTORIA is browser-based, the only limitations are the browser and browser version a user has. The site has been tested to work on Chrome versions 85+, Firefox versions 77+, and MS Edge versions 89+, as of May 10, 2021. 

\section{Results}
\label{sec:results}

\subsection{Supported file formats}
The viewer accepts file types of \textit{.egsphant}, \textit{.3ddose}, DICOM CT Image, DICOM Radiotherapy (RT) Dose, and DICOM RT Structure Set \cite{DICOM_2021}. The files can be loaded with drag and drop, or by opening a file manager to navigate through the user directory. Alternatively, there are test files loaded on the site that consist of a \textit{.egsphant} density file, an RT Dose file, and a structure set file of a pediatric patient, for users to test out the website. A timed upload of a 21.7 MB \textit{.egsphant} file with 161$\times$113$\times$85 voxels took 1.4 s to load, while a batch of 171 DICOM CT files with a total size of 90.4 MB and 512$\times$512$\times$171 voxels took 13.2 s to load, using Google Chrome version 90 on OS X El Capitan with a 2.7 GHz processor speed and 2 cores.

\subsection{User interface}
\begin{figure}
  \centering
  \captionsetup{width=0.9\textwidth}
  \includegraphics[width=0.9\textwidth]{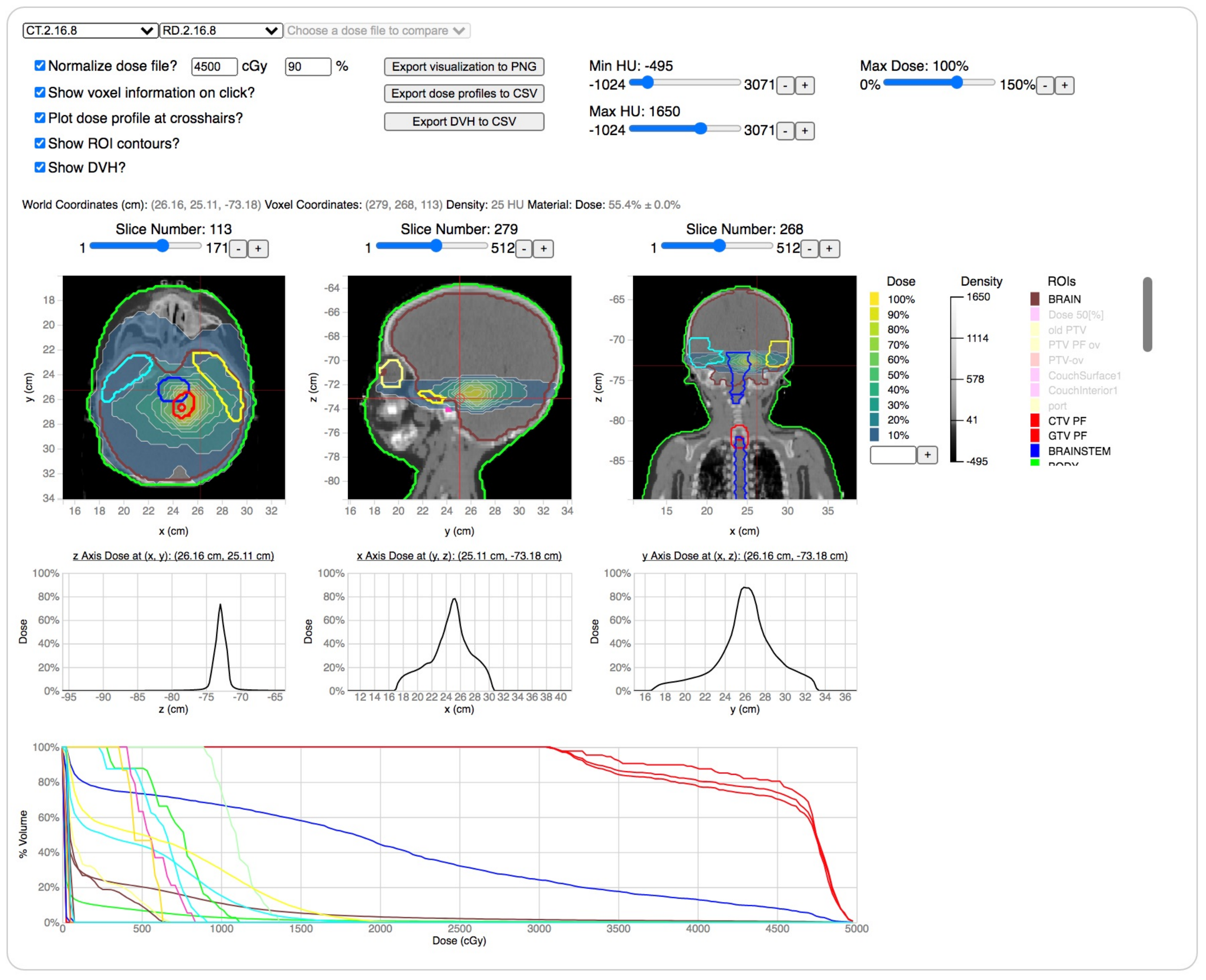}
  \caption{A volume viewer with a CT Image file and RT Dose file loaded and displayed. The window and level of the density file has been adjusted. The dose profile along the red cross-hairs has been plotted. The ROI contours are shown and the corresponding DVH is plotted beneath the dose profiles. The dose has been normalized to 4500 cGy at 90\%, and this is reflected in the DVH x-axis. A few ROI contours have been toggled off.}
  \label{fig:fig1}
\end{figure}

The user interface of VICTORIA needs to display multiple 3D volumes on screen, in each of the three axes. This was accomplished by dividing the website into panes called volume viewers, where each volume viewer can display a density file (either \textit{.egsphant} or CT Image), a dose file (\textit{.3ddose}, RT Dose, or a dose difference), and region of interest (ROI) contours in three panels for the sagittal, coronal, and axial views. The density and dose can be selected using drop down menus at the top of each volume viewer, and the ROI contours are matched by the study unique identifier (UID) or position. 

\subsection{Navigation}
The volumes need to be easy and quick to navigate, without noticeable lag for the best user experience. Each panel can traverse through a view of the volume by moving though its slices. Since the navigation of the slices is based on position rather than voxel number, files that have incompatible voxel dimensions but the same position can still be viewed together. To increase the speed of navigating through a volume, density caching with Web Workers was implemented. Using Google Chrome, navigating each slice for a 21.7 MB \textit{.egsphant} file took 9.2 ms before caching, and 3.1 ms after. For a batch of 171 DICOM CT files with a total size of 90.4 MB, navigating a slice took 45.5 ms before and 14.07 ms after caching, respectively.

\subsection{Data display}
The ability to view the difference between two dose files is a useful feature. This can allow users to compare two different patient plans generated by either a commercial treatment planning system (TPS) or by MC codes (i.e. EGSnrc or TOPAS), to compare doses calculated by a MC code  against a commercial TPS, to compare dose distributions calculated by two difference MC codes, or to compare MC dose distributions calculated with different transport parameters. The dose difference is calculated by dividing each dose matrix by its maximum dose value, then subtracting the second dose matrix from the first, using tri-linear interpolation if the voxel dimensions are different. The panels are navigable by using sliders to move through slice by slice, or by clicking on a point to update the other two panels to the same location. The panels also support panning and zooming. 

The density files are displayed as an image, whereas the dose files are displayed as isodose contours. The dose contours are calculated using the D3 contours function, which returns an array of polygons where the input values are greater than the given thresholds, without any smoothing or interpolation \cite{d3contours_2021}. The default dose contours are from 10\% to 100\% in 10\% increments.  An input box at the bottom of the dose contour legend gives the user the option to input custom dose contours that are not already displayed (e.g., 5\%). A toggle function is also available to show/hide individual dose contours by clicking the corresponding legend cell. The option to plot the dose profiles is also available when a dose file is loaded. The dose profiles contain the dose data along the cross-hairs where the user clicks in the z, x, and y dimensions.

Similar functionality to the dose contours is available for the ROI contours. The structure sets are matched to the volume by the existing density and dose files, rather than the user selecting which structure set to display. They are matched if they have the same study instance UID or if the extent of the density/dose overlap with the ROI contours. The user can toggle on and off which ROI contours are visible in the volume viewer. If there is a dose file loaded, there is also an option to plot a dose volume histogram (DVH). This will go through each ROI contour in the structure set, determine the dose enclosed by each, and calculate the cumulative dose delivered in each portion of the ROI. Again, toggling the ROI legend will toggle the data shown in the DVH plot. 

There are various options the user can make with the check-boxes at the top of each volume viewer. The user can normalize the dose files from cGy to percent, or vice versa. This changes the scale for the DVH. The normalization from percent to cGy is mandatory if the user would like to plot a DVH while using a dose file with relative doses. The voxel information of the position of the panel click can also be displayed. This will show the voxel number, absolute position, density, dose, and material of the position where the user clicks. Another checkbox can plot the dose profile along the cross-hairs of where a user clicks if a dose volume is currently displayed in the viewer. If a set of ROI contours has been uploaded, a checkbox will be available to show or hide them. If ROI contours and dose data are plotted, an option to plot a DVH is also available.

The sliders on the upper right of each volume viewer control different display options. The density sliders can be used to adjust the window and level of the density plots. The max dose slider adjusts the maximum dose for plotting purposes (i.e. if the max dose is 100 cGy, and the max dose slider is set to 80\%, the plot will now be adjusted so the dose at 100\% is 80 cGy, adjusting the dose contours levels). If a dose comparison is plotted, the dose comparison normalization slider will behave similarly, adjusting the normalization factor of the second dose volume that is being compared, from 0.5 to 1.5.

\subsection{Export formats}
VICTORIA offers a range of data export formats for plot data and visualizations. Exporting plot data to CSV allows for analysis in spreadsheet software. If the dose profiles are displayed, the data can be downloaded by selecting the 'Export dose profiles to CSV' button. The positions and dose values of the three dose profiles are then downloaded as a CSV file. The 'Export DVH to CSV' button works similarly, exporting the percent of the ROI volume with the dose value for each of the ROIs. To export the visualization as an image, the 'Export visualization to PNG' button can be used to download a screenshot of the current viewer.

\section{Discussion}
\label{sec:discussion}
VICTORIA was evaluated against some existing free web-based DICOM viewers based on format support and performance. Viewers were only tested if they required no downloads, required no account creation, did all processing client-side, and were free to use. The Imaios DICOM Viewer was evaluated \cite{imaios}. It was very fast and easy to use but did not support the display of DICOM RT Dose nor DICOM RS Structure Set files. Similarly, both the U Dicom Viewer and the Web Dicom Viewer were evaluated as quick and simple DICOM viewing tools that can view remote files via URL, however, both viewers only had support for DICOM CT Image files \cite{udv_2020, dwv_2021}. The Rocket Viewer was another medical image viewing website, with support for a large variety of file types however, it did not support the DICOM RT Dose and RT Structure Set file types, and slices had to be clicked through to navigate \cite{rocket_2018}. The majority of the online DICOM viewers were fast and easy to use, and many had features such as the ability to draw on lines, make measurements, and import remote files. However, none of the tested viewers could upload all of DICOM CT Image, RT Dose, and RT Structure Set files. Additionally, websites that could upload both CT Image and RT Dose files could not overlay them. Furthermore, they did not have features such as dose profiles nor DVH plotting. Currently, VICTORIA is the only online viewer able to read and display the \textit{.egsphant} and \textit{.3ddose} file types.

The VICTORIA viewer is in the early stages of development and evaluation and could be further enhanced. There are many data manipulation features and user interface design improvements to turn it into a robust viewer. Inspired by other viewers, a desirable feature would be the ability to import remote files hosted on DropBox or GoogleDrive. A library to convert a list of 2D polygons to a 3D object should be included to build smoother ROI contours. Another important feature would be speeding up the upload of large files. Some smaller improvements to consider adding to VICTORIA would include a drop down menu to change the colour mapping for the dose contours, a dose contour opacity slider, the ability to close a volume viewer, allowing folder uploads, and an option to plot the density in the dose profile plots. In you are interested in contributing to this project, please head over to our github page \footnote{\url{https://github.com/ellieb/EGSnrc/tree/DICOM-viewer/HEN_HOUSE/gui/dose-viewer}} and submit a pull request. Any questions about the project can be directed to \href{mailto:ebadun@uvic.ca}{ebadun@uvic.ca}.

\section{Conclusion}
\label{sec:conclusion}
VICTORIA is a convenient web-based Monte Carlo and DICOM dose visualization and comparison tool. It is unique in the fact that it is a web-based tool that supports \textit{.3ddose} and \textit{.egsphant} file types. The viewer can be accessed at \url{http://web.uvic.ca/~bazalova/VICTORIA/} \cite{victoria_2021}.

\bibliographystyle{unsrt}

\end{document}